# A cry for help: Early detection of brain injury in newborns

Charles C. Onu[1,2,3]*, Samantha Latremouille[1], Arsenii Gorin[1], Junhao Wang[1], Innocent Udeogu[1], Uchenna Ekwochi[4], Peter O. Ubuane[5], Omolara A. Kehinde[5,6], Muhammad A. Salisu[5,6], Datonye Briggs[7], Yoshua Bengio[2,8], Doina Precup[2,3,9]

[1]Ubenwa Health, Montréal, Canada
[2]Mila-Québec AI Institute, Montréal, Canada
[3]McGill University, Montréal, Canada
[4]Enugu State University Teaching Hospital, Enugu, Nigeria
[5]Lagos State University Teaching Hospital, Lagos, Nigeria
[6]Lagos State University College of Medicine, Lagos, Nigeria
[7]Rivers State University Teaching Hospital, Port Harcourt, Nigeria
[8]Université de Montréal, Montréal, Canada
[9]DeepMind, Montréal, Canada

*Correspondence and requests for materials should be addressed to cc.onu@ubenwa.ai

# Abstract

Since the 1960s, neonatal clinicians have known that newborns suffering from certain neurological conditions exhibit altered crying patterns such as the high-pitched cry in birth asphyxia[1,2]. Despite an annual burden of over 1.5 million infant deaths and disabilities[3,4], early detection of neonatal brain injuries due to asphyxia remains a challenge, particularly in developing countries where the majority of births are not attended by a trained physician[5]. Here we report on the first inter-continental clinical study to demonstrate that neonatal brain injury can be reliably determined from recorded infant cries using an AI algorithm we call *Roseline*. Previous and recent work has been limited by the lack of a large, high-quality clinical database of cry recordings, constraining the application of state-of-the-art machine learning. We develop a new training methodology for audio-based pathology detection models and evaluate this system on a large database of newborn cry sounds acquired from geographically diverse settings – 5 hospitals across 3 continents. Our system extracts interpretable acoustic biomarkers that support clinical decisions and is able to accurately detect neurological injury from newborns' cries with an AUC of 92.5% (88.7% sensitivity at 80% specificity). Cry-based neurological monitoring opens the door for low-cost, easy-to-use, non-invasive and contact-free screening of at-risk babies, especially when integrated into simple devices like smartphones or neonatal ICU monitors. This would provide a reliable tool where there are no alternatives, but also curtail the need to regularly exert newborns to physically-exhausting or radiation-exposing assessments such as brain CT scans. This work sets the stage for embracing the infant cry as a vital sign and indicates the potential of AI-driven sound monitoring for the future of affordable healthcare.

# Main

Crying is the first language of human infants. It is common knowledge that babies cry to express their needs, such as sleep, food or a diaper change. It is, however, much less known that a baby's cry can contain signs of health issues. Clinical scientists in the 1960s to 1980s first drew correlations between patterns of crying and medical conditions, especially neurological and respiratory issues[6]. Through the use of spectrographic analysis, they observed notable differences between the cries of healthy babies and of those impacted by conditions like meningitis, encephalitis, hydrocephalus, Down's syndrome and others[7–9]. The earlier research was often limited by several factors of the times, including the use of analog recordings that could not be flexibly and repeatedly manipulated, storage limitations which meant raw recording files could not be saved, and far less maturity of automated algorithms for signal processing and machine learning. More recent efforts such as the availability of the Baby Chillanto Database[10], helped renew interest in infant cry research[11–13], but given the small number of subjects (<100), its usefulness in developing real world solutions is limited.

We report the results of a first-of-its-kind multicenter, cross-continental clinical study aimed at understanding patterns of crying in infants and characterizing the relationship between these patterns and neurological health. In the study, involving one low-income (Nigeria), one middle-income (Brazil) and one high-income country (Canada), we prospectively enrolled a total of 2,631 patients across 5 hospitals.

Specifically, we recruited patients impacted by perinatal asphyxia, a condition where the newborn lacks appropriate oxygenation at birth for a variety of reasons[14]. Depending on the duration and severity of the hypoxic event, it can damage the brain to varying degrees, causing a condition known as neonatal encephalopathy (or brain injury). Globally, asphyxia is the second most common cause of newborn fatality[3], accounting for half a million deaths and over a million disabilities in those who survive, including cerebral palsy and other neurodevelopmental impairments[4,14]. Early identification of the signs of evolving brain injury is critical in order to start life-saving interventions, such as therapeutic hypothermia and pharmacological support. However, such neurological examinations require trained personnel; a problem for low- and middle-income countries where the majority of births are not attended by skilled personnel. Apgar scoring, a rapid physical examination of the newborn immediately after birth to evaluate their health during post-birth transition and resuscitation, is typically the only available option. Unfortunately, Apgar scores have not been reliable for identifying those who will progress to brain injury[15–17]. In particular, as Apgar scoring is only done at the first 1, 5, and 10 minutes of life, it does not take into account the infant's cardiorespiratory health after this time. Automated infant cry analysis offers a simple, non-invasive and contact-free way of assessing the neurologic state of an infant in the hours after delivery.

We developed a new training methodology for audio-based pathology detection models and applied it to cry analysis for identifying signs of neurological injury. Using only cry sounds at birth, our system, depicted in Fig. 1a, which we name *Roseline* (Reduction Of Self-supervised Entropy to Learn and Infer Neonatal Encephalopathy), is able to identify evolving neurological injury with an AUC of 92.5% and 88.7 sensitivity at 80% specificity (Fig. 1b). Our training methodology is a 3-step process which employs self-supervised learning to extract a rich representation from large databases of diverse audio recordings including adult speech, music, animal and urban sounds, then adapts and transfers the representation as the foundation for training infant cry analysis models. Beyond developing predictive models, we investigated and identified key audio features as "acoustic biomarkers". These biomarkers extracted from the infant cry sounds effectively measure physiological parameters in the infant's body, combining them to offer a clinical decision process that is not only accurate but also interpretable to physicians (Fig 1c).

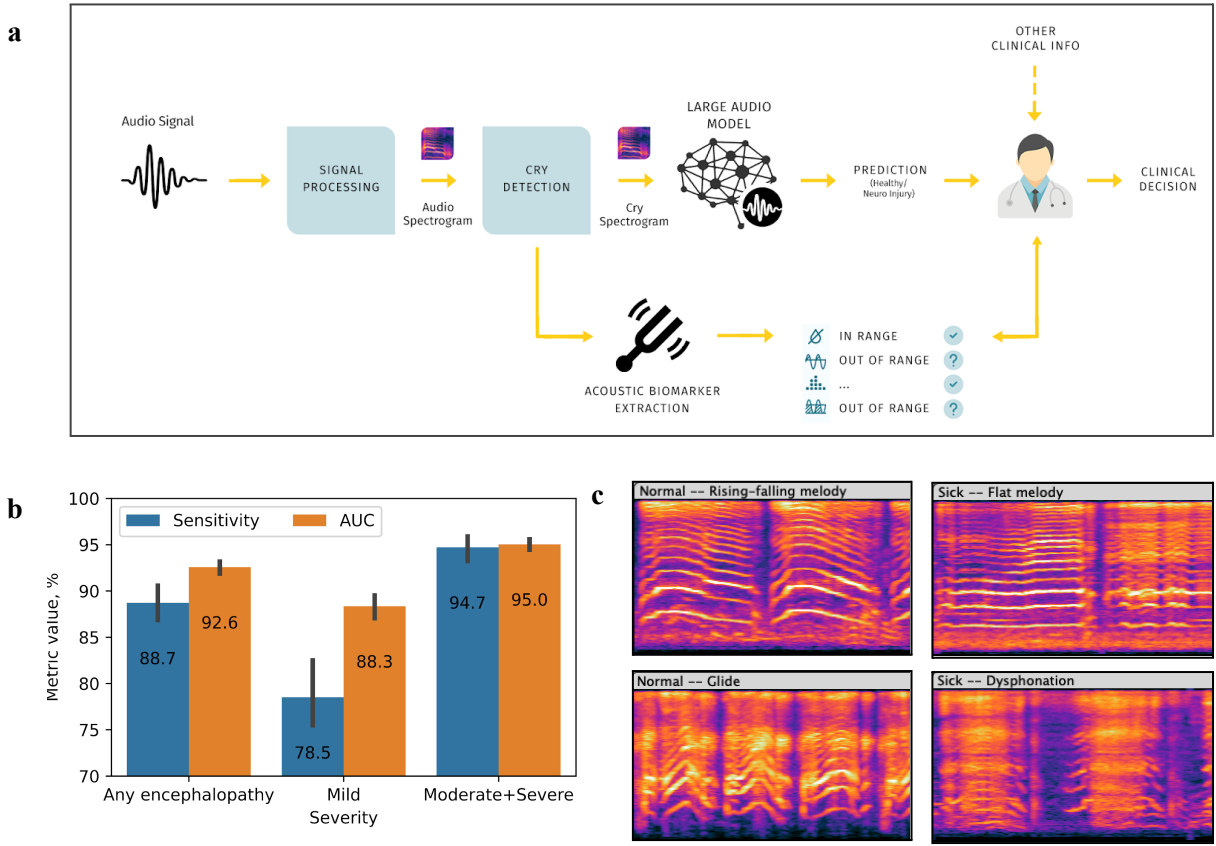

*Fig 1. Roseline - AI-based neurological screening supported with acoustic biomarkers. **a,*** *The system takes cry signal as an input and provides clinicians with two types of information - neuro injury predicted by neural network-based large audio model (LAM) and acoustic biomarkers that summarize cry characteristics. Both LAM and biomarker analyze isolated cry signals extracted by an automatic cry activity detector to minimize the impact of external noises.* ***b.*** *For reliable LAM evaluation, the model was trained 10 times with different random seeds and tested on a held-out set. The figure reports average sensitivity at 80% specificity and AUC with standard error. As expected, the mild disease is harder to detect (88.3% AUC) compared to moderate and severe cases (95.0% AUC);* ***c.*** *With the black-box nature of neural network models, it is important to provide clinicians with as much information about model predictions as possible. We analyzed more than a hundred signal processing-based acoustic biomarkers and selected a compact set of features that consistently correlate with neurological injury or its absence.*

# A prospective, multi-center, intercontinental clinical study to characterize the infant cry

Over a period of 3 years, we conducted prospective clinical studies at 5 health networks in Brazil, Canada and Nigeria, namely Santa Casa de Misericordia de Sao Paulo (SCDM), McGill University Health Centre (MUHC)[1], Enugu State University Teaching Hospital (ESUTH), Rivers State University Teaching Hospital (RSUTH) and Lagos State University Teaching Hospital (LASUTH).

The study, dubbed *Ubenwa* (meaning "cry of a baby" in Igbo language), enrolled a total of 2,631 term newborns, i.e., of at least 36 weeks of gestational age. Newborns belonged to one of two cohorts: (a) "asphyxia cohort", that is those who were admitted to the hospitals' neonatal intensive care units (NICU) with a history of a hypoxic insult, including sentinel events, Apgar scores, resuscitation requirements, and/or blood gasses where available; (b) "healthy cohort", that is patients who had no evidence of a hypoxic insult, typically recruited from the normal newborn nurseries. Patients received a neurological assessment at birth (or admission) and at discharge by a clinician using a modified Sarnat scoring system[18,19] – assigning the level of injury as normal, mild, moderate or severe. At the time of Sarnat assessments, a recording of the newborn's unelicited cry was obtained using an in-house developed mobile application (see Extended Data Fig. 1) on a Samsung A10 smartphone held at 10-15 cm from the newborn's mouth. Figure 2 summarizes the study protocol and data distribution across countries. In this work, we set out to utilize a cry sample taken after birth ("*birth assessment*" in Fig 1a) in identifying the presence and severity of evolving brain injury, sequel to perinatal asphyxia.

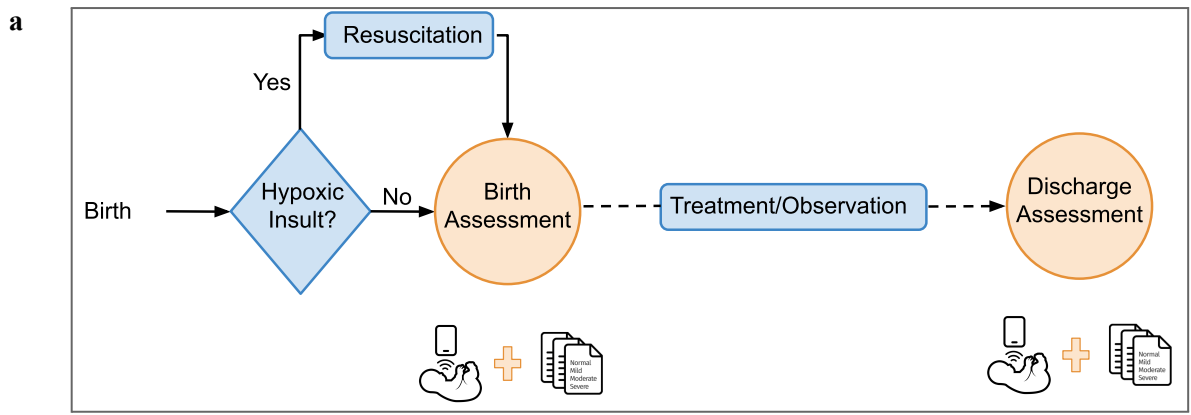

---

[1] The Ubenwa study at the MUHC has closed. Other sites continue to follow patients longitudinally to monitor future neurodevelopmental outcomes.

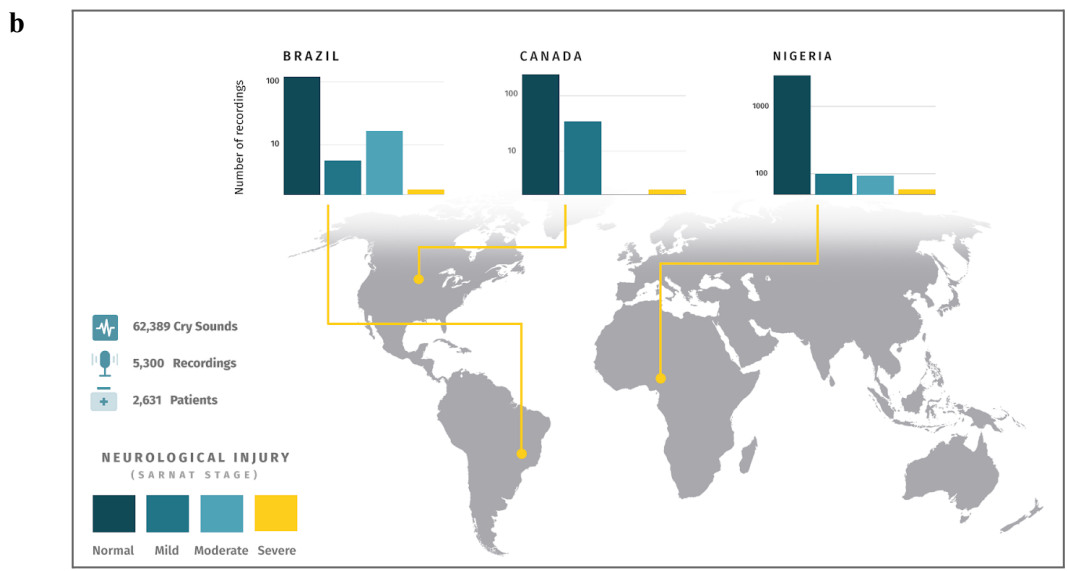

*Fig 2. A multi-center, cross-continental clinical study. **a**,* Infant cry recordings are taken at 2 time points – birth at birth and at discharge. Each time a 30-sec cry sample is obtained and a Sarnat exam conducted by a neonatologist. ***b**,* Recordings are collected from healthy controls and babies diagnosed with neurological injury. ***b**,* Five hospitals in 3 countries participated in the study with patients categorized as either normal or with mild, moderate and severe neurological injury.

# Predicting neurological injury from cry sounds

Since the advent of foundation models in large language models (LLMs) which are now so successful in multiple text processing tasks, the power and flexibility of models that are pre-trained in a self-supervised manner has become evident[20,21]. In a similar way, we develop a large audio model (LAM), enabling the foundational model to acquire a strong representation that can be used for many downstream tasks including cry analysis.

Specifically, we designed a 3-stage training methodology: (1) **pre-training** a foundation model on a massive audio data set (2) **adaptation** using an unlabelled subset of our cry database and (3) **fine-tuning** using the clinically-annotated cries. This approach is illustrated in Figure 3a. The base model (encoder) is a 76M-parameter convolutional neural network that has demonstrated strong performance across a wide range of audio processing tasks including emotion detection[22], sound event detection[23], and music classification[24]. Our model consumes cry signals in the form of audio spectrograms, a time-frequency representation computed using the short-time fourier transform. Spectrograms are widely adopted in ML-based audio classification as they allow efficient application of computer vision methods to sound.

In the pre-training phase (stage 1), we aim to build high-quality audio representations leveraging large amounts of open-source audio data. We achieve this in a self-supervised manner without data labels, employing the similarity-based contrastive learning method called SimCLR[25]. Precisely we train our model to recognize if spectrograms of two short audio clips corrupted with random time-frequency masks were extracted from the same audio recording. For pre-training, we use the well-known VGG sound[26] database, which contains 550 hours of over 300 audio classes. In experiments, we found the use of self-supervised pre-training to be critical, contributing a 3.8% gain in downstream performance (AUC) over a model that was pre-trained in a conventional supervised manner using labels (Fig. 3b).

In the adaptation phase (stage 2), we specialize our audio representation model from the domain of all sounds to the domain of cry sounds. The data used here comes from unlabeled samples from the Ubenwa study, typically samples with incomplete data collection, such as missing Sarnat annotations and discharge recordings. The SimCLR framework is again employed here to update the learned weights of our LAM, in a self-supervised manner. The adaptation step is important as it tailors the foundation model's parameters to the uniqueness of infant cry recordings. Our adapted model surpasses the downstream AUC of a non-adapted model in predicting neurological injury by 2.9%.

In the third and final stage, we fine-tune the LAM using the carefully annotated cry samples from the Ubenwa labeled cry database, which is divided into a training, validation and test portion. In this stage the model is trained to classify cry sounds based on the patient's state – healthy or neuro injury. The 10-fold cross-validation approach is used to select the model's hyperparameters (see Methods and Extended Data Fig. 2). The model is then evaluated using the left-aside, test portion of the labeled cry database. The results are summarized in Figure 3b. The resulting LAM achieves 92.5% AUC in detecting neurological injury from cry sounds, with a sensitivity of 88.7% at 80% specificity. Detailed technical information on the 3 stages can be found in the Methods section of this article.

While examining the breakdown of model accuracy by the different levels of neurological injury (Figure 1b), we observe that the model more easily identifies patients with moderate and severe neurological injury (95.0% AUC) than the milds (88.3% AUC). As part of future work it will be important to find and flag more signs of mild injury. Close monitoring of mild patients is crucial given limited evidence available for their appropriate treatment options, while they remain at risk of exacerbation of their condition[27].

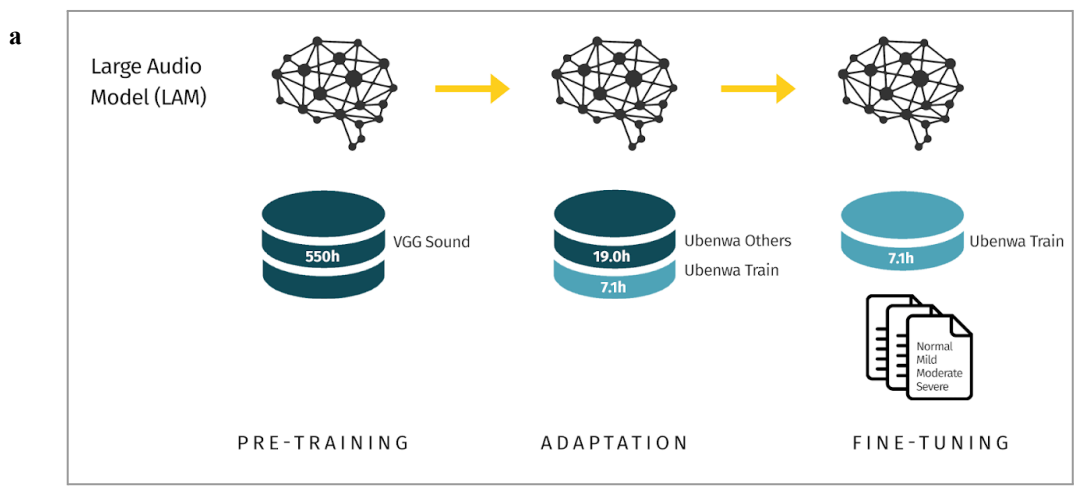

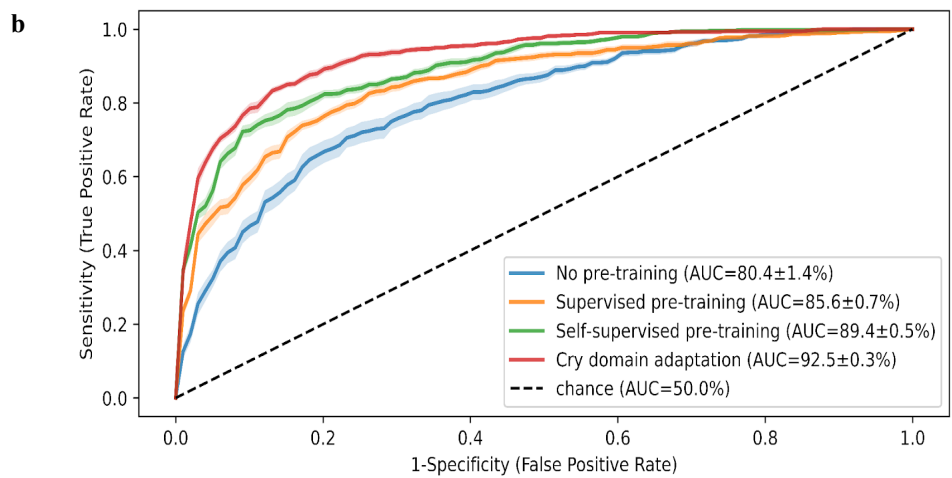

*Fig 3: A machine learning model for audio-based pathology detection*. **a,** The model for identifying neurological injury from cry sounds is trained in 3 stages: pre-training, adaptation and fine-tuning. At the core is a large audio model (LAM), a convolutional neural network pre-trained on a broad public audio classification dataset called VGGSound, which consists of about 550 hours of data. This generic pre-trained audio model is further adapted to deal specifically with cry sounds by means of a self-supervised learning algorithm, which only uses cry sounds and not the associated labels. A smaller portion of clinically annotated cry sounds is finally used to "fine-tune" the model for the final task of asphyxia prediction by using the associated label - Sarnat scores associated with each audio. **b,** Average AUC across 10 randomized runs is used to compare models. The best model (red line) that is both pre-trained on general audio and adapted to cries yields a 92.5% AUC score.

# Acoustic biomarkers for explainability and clinical decision support

Given the rich database acquired, we study the opportunity for *acoustic biomarkers* of the infant cry to deepen our understanding of how pathology alters cry patterns and as a means of model explainability. Such explainability has immense value in AI-based clinical decision support as it advances science in an era of black box predictors, and keeps control in the hands of physicians thereby providing an opportunity for safe and robust decision-making.

To develop these acoustic biomarkers, we studied two kinds of features: *generic voice features* and *cry-specific biomarkers*. Generic voice features include measures commonly used in audio analysis such as fundamental frequency, resonance frequencies and cepstral coefficients. Cry-specific biomarkers are higher-level features that measure specific aspects of the infant's physiology. For instance, dysphonation (noisy inharmonic cry) indicates unstable respiratory control in the lower vocal tract. See an overview of 8 cry-specific biomarkers in Figure 4a. Some of these cry-specific biomarkers were first introduced by Truby, H. et al in 1960s[28]. We refined the definitions, and developed signal-processing algorithms leveraging our much larger database.

We employed a 3-step approach to develop and analyze a compact set of specialized acoustic biomarkers of neurological injury. First, a complete set of audio features is extracted for each recording. It consists of 88 temporal derivatives of generic voice features and 26 derivatives of cry-specific biomarkers (see details in the Methods section). Next, the full set is reduced by selecting a smaller number of highly predictive biomarkers. The Pearson's correlation coefficient (PCC) of each feature was computed with the outcome (healthy or neuro injury) to determine the directionality (positive or negative) of each feature with respect to the outcome. This was performed on a per-hospital basis, and a feature was *selected* if its PCC has the same directionality across all hospitals, indicating that this feature has a consistent, non-spurious relationship with the infant's health state. Finally, in order to validate the utility of the selected features, a linear classifier was trained to predict neurological injury. Linear classifiers are useful for this kind of analysis as they quantify the relative contribution of each feature as percentage weights in the model (see Methods for more details).

Using this procedure, we identified the top 18 audio features that have a consistent correlation with neurological injury across hospitals. We call these *acoustic biomarkers* of the infant cry. They include, for example: *dysphonation* which measures non-harmonic pitch profiles or vocal cord vibrations; *melody type* which measures the trend of the pitch (rising, falling, flat); and *glide* which measures rapid changes in pitch. We found that 12.1% sick babies had dysphonation in more than half of their cry expirations, in contrast to only 7.7% of healthy babies. 73.8% of sick babies had flat melody type in more than half of their cry expirations, in contrast to only 62.4% of healthy babies. In Figure 4b, we summarize some of these findings as box plots.

a

**Time-based biomarkers**

**Cry Unit Duration**
Time from the onset to offset of an expiratory cry utterance.

**Pause Duration**
Time between the offset of one expiratory utterance and the onset of the next.

**Pitch-based biomarkers**

**Hyperphonation**
High-pitched cry-unit segment during an expiratory utterance with a fundamental frequency typically higher than 1000 Hz.

**Dysphonation**
Noisy, turbulent, or inharmonic cry-unit segment during an expiratory utterance.

**Glide**
Rapid change of fundamental frequency observed in an expiratory phonation, usually of short duration.

**Vibrato**
Rapid oscillations of fundamental frequency within one expiratory utterance.

**Melody type**
Fundamental frequency variations within one expiratory utterance, defined in five categories: falling, rising-falling, rising, falling-rising, and flat.

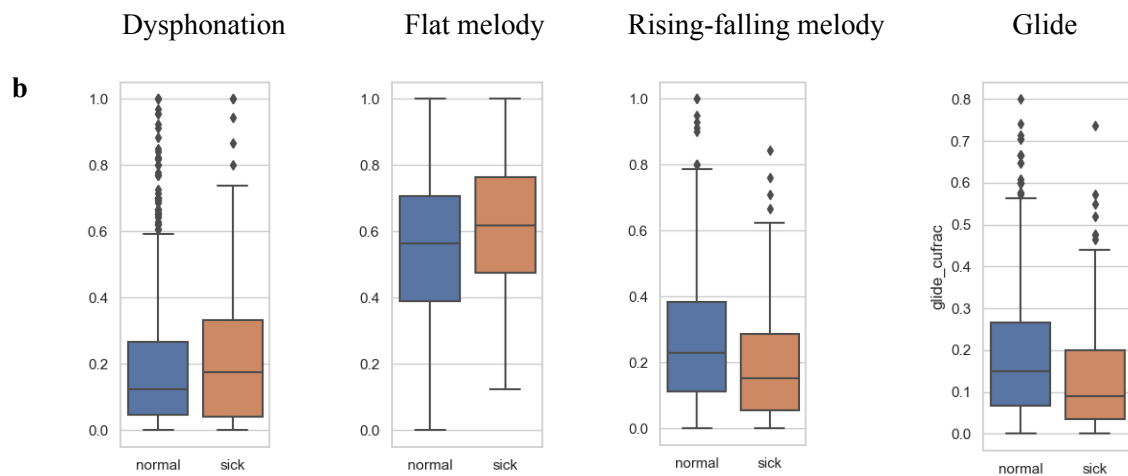

**Fig 4 Analysis and development of acoustic biomarkers from the infant cry. a,** The 8 acoustic biomarkers developed and their definitions, **b,** Some of the best-performing biomarkers show significant differences across sick and healthy patient recordings from our clinical database. For example, dysphonation and flat melody type are two biomarkers that consistently correlate with neuro injury across all hospitals. In contrast, rising and falling pitch melody type and pitch glide were more frequently encountered in the cries of healthy babies.

# Discussion

In this article, we present *Roseline*, a deep learning system for identifying newborns at risk of brain injury solely using recordings of their cry sounds. To develop this system, we acquired a large and geographically diverse clinical database of cry recordings and developed a new training methodology for audio-based pathology detection models. The development and validation of acoustic biomarkers of the infant cry in this work also marks an important step towards understanding the impact of pathology on patterns of crying and connecting it to the physiology of cry production. This finding makes it possible to not only develop new tools but to also expand our scientific understanding of infant crying. This work was motivated by the high rate of newborn casualty (death and disability) in low-resource settings. Every year over 1.5 million newborns die or are disabled for life due to brain damage from birth asphyxia. With our methodology, which facilitates detection at the earliest point of life, we believe this number can be greatly reduced.

Precisely, the use of the cry as a diagnostic input opens up a world of opportunities for easy-to-use, non-invasive, contact-free monitoring of neonates. In particular, within developing countries and all medically underserved areas, we can bring a neurological examination performed by specialized medical personnel into the hands of any birth attendant. It would effectively convert low-cost devices like smartphones and wearables into medical devices, drastically reducing the cost of access. Furthermore, in patients requiring close monitoring in the neonatal intensive care unit, cry analysis for accurately tracking neurological health could limit or prioritize the need for additional costly tests, such as electroencephalography (EEG) and brain imaging.

An important limitation of this algorithm lies in the identification of the Mild levels of encephalopathy. However, this also relates to an important limitation of the Sarnat exam with respect to the Milds, as the criteria categorizing a patient as a Mild is the most broad, where any abnormal finding in any of the evaluated categories can be considered Mild unless fulfilling the more strict and severe definition of Moderate or Severe. In fact, a significant proportion of Milds go on to have short-term and long-term neurodevelopmental abnormalities, reportedly ranging from 16-52% depending on the definition of Mild and the treatments received[19,29,30]. This range highlights the importance of finding biomarkers to identify the proportion of Mild patients that go on to have significant brain injury, such that appropriate interventions can take place[29]. In our research, neurodevelopmental follow-ups are currently underway, and this will allow for the identification and training of the models on potentially different "levels" of Milds, as well as for the prediction of long-term neurological injury in all levels of encephalopathy. Thus, cry analysis may play a critical role in identifying the most at-risk infants that could benefit from specific treatments.

There are social and ethical considerations to be taken into account when deploying any AI system, and ours is no exception. Adequate care must be taken for user privacy since the voice of a baby may be considered biometric data. Recently, we open-sourced a portion of the Ubenwa database for a machine learning competition on speaker verification using infant cry sounds[31]. Though this revealed that it is very challenging to identify a baby from their cry, we expect that with time and more research progress, it may become possible. The impact of noise on our model may also be further studied. Recordings were made in real hospital environments, so it was necessary to use a cry detection algorithm to accurately isolate noises from cry sounds before analysis. In deployment, it may be necessary to further study the impact of background noise on model performance.

We expect that this is only the beginning and further progress will be made in developing the infant cry as a vital sign for not only neurological injury but also a range of other conditions. We have ongoing collaborations to investigate cry as a diagnostic in other medical conditions as well as the prognostic value of cry analysis. A percentage of patients who suffer asphyxia will have long-term neurodevelopmental issues. As a prognostic, cry analysis could help with early identification of patients who are at risk of issues such as learning difficulties, autism, speech delays and others, increasing the chance that infants receive help that enables them to survive and thrive.

# References

1. Michelsson, K. Cry analysis of symptomless low birth weight neonates and of asphyxiated newborn infants. *Acta Paediatr.* **60**, 9–45 (1971).

2. Partanen, T. J. *et al.* Auditory identification of pain cry signals of young infants in pathological conditions and its sound spectrographic basis. *Ann. Paediatr. Fenn.* **13**, 56–63 (1967).

3. Lawn, J. E., Cousens, S. & Zupan, J. 4 million neonatal deaths: When? Where? Why? *The Lancet* **365**, 891–900 (2005).

4. Newborn care. *UNICEF DATA* https://data.unicef.org/topic/maternal-health/newborn-care/.

5. Montagu, D., Yamey, G., Visconti, A., Harding, A. & Yoong, J. Where do poor women in developing countries give birth? A multi-country analysis of demographic and health survey data. *PloS One* **6**, e17155 (2011).

6. Wasz-Höckert, O., Michelsson, K. & Lind, J. Twenty-Five Years of Scandinavian Cry Research. in *Infant Crying* (eds. Lester, B. M. & Zachariah Boukydis, C. F.) 83–104 (Springer US, 1985). doi:10.1007/978-1-4613-2381-5_4.

7. Saraswathy, J., Hariharan, M., Yaacob, S. & Khairunizam, W. Automatic classification of infant cry: A review. in *2012 International Conference on Biomedical Engineering (ICoBE)* 543–548 (2012). doi:10.1109/ICoBE.2012.6179077.

8. Kheddache, Y. & Tadj, C. Resonance Frequencies Behavior in Pathologic Cries of Newborns. *J. Voice* **29**, 1–12 (2015).

9. Jeyaraman, S. *et al.* A review: survey on automatic infant cry analysis and classification. *Health Technol.* **8**, 391–404 (2018).

10. Reyes-Galaviz, O. F. & Reyes-Garcia, C. A. A System for the Processing of Infant Cry to Recognize Pathologies in Recently Born Babies with Neural Networks. in *SPECOM'2004: 9th Conference on Speech and Computer* 552–557 (2004).

11. Onu, C. C., Lebensold, J., Hamilton, W. L. & Precup, D. Neural Transfer Learning for Cry-Based

Diagnosis of Perinatal Asphyxia. in *Interspeech 2019* 3053–3057 (ISCA, 2019). doi:10.21437/Interspeech.2019-2340.

12. Patil, H. A., Patil, A. T. & Kachhi, A. Constant Q Cepstral coefficients for classification of normal vs. Pathological infant cry. in *ICASSP 2022 - 2022 IEEE International Conference on Acoustics, Speech and Signal Processing (ICASSP)* 7392–7396 (2022). doi:10.1109/ICASSP43922.2022.9746946.

13. Kachhi, A., Chaturvedi, S., Patil, H. A. & Singh, D. K. Data Augmentation for Infant Cry Classification. in *2022 13th International Symposium on Chinese Spoken Language Processing (ISCSLP)* 433–437 (2022). doi:10.1109/ISCSLP57327.2022.10037931.

14. Antonucci, R., Porcella, A. & Pilloni, M. D. Perinatal asphyxia in the term newborn. *J. Pediatr. Neonatal Individ. Med. JPNIM* **3**, e030269–e030269 (2014).

15. Marrin, M. & Paes, B. Birth asphyxia: does the Apgar score have diagnostic value? *Obstet. Gynecol.* **72**, 120–3 (1988).

16. Ellis, M., Manandhar, N., Manandhar, D. S. & deL Costello, A. M. An Apgar score of three or less at one minute is not diagnostic of birth asphyxia but is a useful screening test for neonatal encephalopathy. *Indian Pediatr.* **35**, 415–421 (1998).

17. Aliyu, I., Teslim, L. & Onankpa, B. Hypoxic-ischemic encephalopathy and the Apgar scoring system: The experience in a resource-limited setting. *J. Clin. Sci.* **15**, 18 (2018).

18. Sarnat, H. B. & Sarnat, M. S. Neonatal Encephalopathy Following Fetal Distress: A Clinical and Electroencephalographic Study. *Arch. Neurol.* **33**, 696–705 (1976).

19. Prempunpong, C. *et al.* Prospective research on infants with mild encephalopathy: the PRIME study. *J. Perinatol. Off. J. Calif. Perinat. Assoc.* **38**, 80–85 (2018).

20. Bommasani, R. et al. On the Opportunities and Risks of Foundation Models. ArXiv (2021).

21. Brown, T. *et al.* Language Models are Few-Shot Learners. in *NIPS* vol. 33 1877–1901 (Curran Associates, Inc., 2020).

22. Triantafyllopoulos, A. & Schuller, B. W. The Role of Task and Acoustic Similarity in Audio Transfer

Learning: Insights from the Speech Emotion Recognition Case. in *ICASSP 2021* 7268–7272 (2021). doi:10.1109/ICASSP39728.2021.9414896.

23. Kong, Q. *et al.* PANNs: Large-Scale pre-trained Audio Neural Networks for Audio Pattern Recognition. *IEEE ACM Trans Audio Speech Lang Process* **28**, 2880–2894 (2020).

24. Lee, J., Park, J., Kim, K. L. & Nam, J. SampleCNN: End-to-end deep convolutional neural networks using very small filters for music classification. *Appl. Sci.* **8**, 150 (2018).

25. Chen, T., Kornblith, S., Norouzi, M. & Hinton, G. E. A Simple Framework for Contrastive Learning of Visual Representations. in *ICML 2020* vol. 119 1597–1607 (PMLR, 2020).

26. Chen, H., Xie, W., Vedaldi, A. & Zisserman, A. VGGSound: A Large-Scale Audio-Visual Dataset. in *ICASSP 2020* 721–725 (IEEE, 2020). doi:10.1109/ICASSP40776.2020.9053174.

27. Kariholu, U. *et al.* Therapeutic hypothermia for mild neonatal encephalopathy: a systematic review and meta-analysis. *Arch. Dis. Child. - Fetal Neonatal Ed.* **105**, 225–228 (2020).

28. Truby, H. M. & Lind, J. Cry Sounds of the Newborn Infant. *Acta Paediatr.* **54**, 8–59 (1965).

29. Chalak, L., Latremouille, S., Mir, I., Sánchez, P. J. & Sant'Anna, G. A review of the conundrum of mild hypoxic-ischemic encephalopathy: Current challenges and moving forward. *Early Hum. Dev.* **120**, 88–94 (2018).

30. Chalak, L. F. *et al.* Prospective research in infants with mild encephalopathy identified in the first six hours of life: neurodevelopmental outcomes at 18-22 months. *Pediatr. Res.* **84**, 861–868 (2018).

31. Budaghyan, D., Onu, C. C., Gorin, A., Subakan, C. & Precup, D. CryCeleb: A Speaker Verification Dataset Based on Infant Cry Sounds. Preprint at https://doi.org/10.48550/arXiv.2305.00969 (2023).

# Methods

## Training datasets

### Cry database

The dataset of cries and associated clinical information was collected by five hospitals between 2020 and 2023. The overall raw database consists of 4,312 recordings from 2,631 patients. In most cases, one recording is done right after birth and one before discharge (one or more days after birth). In total, 344 infants were admitted with symptoms of neurological injury measured by the modified Sarnat[18,19]. In our study, for the experiments that require labels (ML training, biomarker tuning, etc) we use a subset of birth recordings.

### Curated labeled subset of birth recordings for model development

To ensure the high quality of data used in model development, a number of data cleaning steps were done. First, we manually segmented the cry expiratory sounds in the audio files recorded prior to November 2022. We then only kept audio files that had at least 3 seconds of cry sound.

In addition, the annotations were cross-referenced with independently collected clinical records (data collection forms) and a number of recordings were removed as potentially mislabeled resulting in 2,174 recordings with 17 hours of cry signal in total. Finally, we excluded discharge recordings as we focus on birth screening in this study. The resulting curated labeled subset of birth recordings consists of 1,108 audio samples - 959 recordings of controls and 149 diagnosed with neuro injury. This dataset is further subdivided into train, validation and test subsets (see Extended Data Table 1).

### Unlabeled dataset of cry recordings for self-supervised domain adaptation

Data cleaning for self-supervised model adaptation did not have to be cleaned as rigorously as the labeled subset. We took all 4,312 recordings and only excluded recordings of patients that appeared in the validation and test sets. Also, instead of manual segmentation of cry, all audios were processed by an automatic cry activity detector - a small convolutional neural network trained to discriminate cry audio segments. The resulting automatically cleaned unlabeled dataset for self-supervised domain adaptation consists of 3,613 audio recordings.

### VGGSound dataset for model pre-training

VGGSound is a public dataset[26] which consists of 199,176 10-second audio clips from YouTube that are categorized into 308 classes (dog barking, car engine, etc). It includes 667 audio clips of baby crying although not specifically newborns.

# Large audio model

## Model architecture

The model is based on CNN14[23] architecture inspired by popular computer vision VGG architecture[32]. CNN14 demonstrated strong performance across various audio classification tasks when pre-trained on large generic audio data.

The audio signal is first processed using short-time Fourier transform (STFT) to create log Mel filterbanks. The resulting log Mel spectrogram is a 2D representation with 10 millisecond time frames on x-axis and 80 frequency bands on y-axis with values corresponding to the corresponding energies.

The spectrogram is further passed to a neural network with 6 layers, where each block is composed of two convolution layers with 3x3 kernel size, batch normalization and ReLU. The network uses 4-second audio clips during training and arbitrary sequences during inference by means of a global pooling operation at the output of the last convolutional layer.

To summarize, the network contains 76 million parameters which encode arbitrary input audio file into a 2048-dimensional feature vector for classification or contrastive self-supervised learning.

## Pre-training with VGGSound

The initial pre-trained models were provided by Wang, Z. et al[33] with the publicly available training code. Supervised pre-training is done over 200 epochs with stochastic gradient descent with batches of 32 4-second random audio segments of VGGSound. In addition, input spectrograms are corrupted using a popular speech data augmentation method called SpecAugment[34], which randomly masks parts of the spectrograms thus increasing the diversity of samples and model robustness.

Self-supervised pre-training was done by the same authors in a similar manner to supervised pre-training but using SimCLR method. SimCLR is a contrastive self-supervised learning algorithm originally proposed for computer vision[25] that recently demonstrated good results in audio / music classification tasks[33,35]. It maximizes the similarity between two views of the same audio recording. In our implementation, each view is created by randomly extracting a 4-second audio segment and applying SpecAugment to the spectrogram.

## Self-supervised cry domain adaptation

We perform domain adaptation using the technique originally described in[36]. Specifically, starting with the network that is pre-trained with SimCLR on VGGSound dataset, we train 100 epochs using batches of 200 samples where half of the batch is sampled from the unlabeled Ubenwa dataset and another half - randomly sampled from VGGSound. Adding VGGSound to batches can be seen as so called "rehearsal" or "replay" that are common in domain adaptation scenarios (continual learning) for reducing catastrophic forgetting (drop of performance on an initial task when learning new task)[37]. While we are not studying or addressing forgetting in this study, we found in[36] that rehearsal consistently leads to better generalization which is important for supervised transfer learning.

## Supervised transfer learning with target labels

To transfer the pre-trained network to the task of neuro injury prediction we add a randomly initialized classification layer (head) and use labeled data to fully or partially fine-tune the resulting network by minimizing cross-entropy loss.

To ensure that we utilize pre-trained models efficiently, for all models we evaluate three fine-tuning strategies that work best in 10-fold cross-validation along with additional hyperparameter tuning.

Specifically, the following transfer learning strategies are considered:
- Linear probing, where the CNN is used as a feature extractor and only the the linear head is trained. This only requires supervised training of 4,096 parameters and is very stable
- End-to-end fine-tuning where the encoder parameters are also updated
- Linear probing with an update of batch normalization layers of the encoder

In all cases, fine-tuning is done over 50 epochs using Adam optimizer with learning rate reduced by half if validation AUC is not improved over three epochs. Best model parameters are selected from the training loop based on validation AUC that is evaluated every training epoch. For end-to-end fine-tuning, we use a separately tuned learning rate for the encoder parameters along with linear warmup (gradual increase of learning rate) over the first 10 training epochs.

## Hyperparameter tuning

We selected the best fine-tuning strategy and hyperparameters using stratified 10-fold cross-validation. For non-end-to-end approaches, the training was quite stable so the grid search only included a range of learning rates ["5e-5", "1e-4", "5e-4", "1e-3"]. For end-to-end fine-tuning, we search for the best cross-validation AUC in all combinations of learning rates ["1e-4", "5e-4", "1e-3"] for head and ["1e-5", "5e-5", "1e-4", "5e-4", "1e-3"] for the encoder.

For non domain-adapted models, end-to-end fine-tuning always performed best with larger learning rates required for randomly initialized models. For domain-adapted models a much simpler linear probing with batch normalization update consistently performed better.

Extended Data Fig. 2 summarizes cross-validation performance of best hyperparameter settings for random initialization, pre-training and domain adaptation. The results are consistent with the final test set performances summarized in Fig. 3b.

# Statistical-based cry analysis

## Generic voice features as cry biomarkers

Generic voice features used in this study were computed using openSMILE (open-source Speech and Music Interpretation by Large-space Extraction), an open-source toolkit for audio feature extraction and audio classification[38]. Specifically, we make use of the extended Geneva Minimalist Acoustic Parameter set (eGeMAPS), which includes 25 low-level descriptors (LLDs) designed for automatic voice analysis tasks[39]. The LLDs are time sequences computed from the input audio. They are further aggregated into a total of 88 functionals by taking various statistics over the sequence.

The motivation for choosing eGeMAPS for our study is twofold. First, the LLDs in eGeMAPS were selected based on their potential to index physiological properties in human voice, thus making it possible to verify physiological hypotheses against feature-analysis findings. Second, eGeMAPS are widely adopted in voice and audio literature[40–45].

To extract features from an input cry recording, we first collect all expiration cry segments from manually annotated recordings. The expiration segments are concatenated into one audio array and processed by the openSMILE feature extractor. At the output, each cry recording is represented by an 88-dimensional feature vector.

## Cry-specific biomarkers in clinical literature

We refer to cry biomarkers as signal processing-based features that were extensively studied in existing clinical literature on cry-based neurological assessment of newborns.

Specifically, we consider biomarkers based on fundamental frequencies (F0) such as hyperphonation, dysphonation, glide, vibrato, melody types, and time-domain durations: expiratory cry unit duration and pause duration.

**Hyperphonation** is defined as a high-pitched cry-unit segment during an expiratory utterance with a fundamental frequency typically higher than 1000 Hz[6,46,47]. It results from a "falsetto" like vibration pattern of the vocal folds. It has been reported as an indicator of neural constriction of the vocal tract[48] and associated with various pathologies such as laryngomalacia[47], asthma[47], respiratory distress syndrome[47], and prenatal exposure to opiate[49,50], cocaine[51], and alcohol[50].

**Dysphonation** is defined as a noisy, turbulent, or inharmonic cry-unit segment during an expiratory utterance[46,48,52] and has been reported to indicate unstable respiratory control[48]. In existing clinical cry research, it has been associated with depression[48], laryngomalacia[47], congenital heart disease[47], meningitis[47], brain hemorrhage[47], and prenatal exposure to cocaine[50,53], marijuana[54], and alcohol[55].

**Glide** is defined as a rapid change of fundamental frequency observed in an expiratory phonation, usually of short duration[6,52]. In existing clinical studies, glide has been associated with birth asphyxia[1], meningitis[52], and hydrocephalus[52]. It also occurs more frequently in the cry of premature neonates[52].

**Vibrato** is defined to occur when there are at least four rapid up-and-down movements of fundamental frequency within one expiratory utterance[1,52] and has been studied in the context of congenital heart disease[1], deafness[47], and birth asphyxia[1,47].

**Melody type** describes the fundamental frequency variations within one expiratory utterance, defined in five categories: falling, rising-falling, rising, falling-rising, and flat[1,46]. It reflects the trend of fundamental frequency over time. It was reported that a typical cry of the healthy newborn has a falling or rising-falling melody while a significant increase in rising, falling-rising, and flat types of melody was observed in those with central respiratory failure[1,56].

**Cry-unit duration** is defined as the time from the onset to offset of an expiratory utterance[48,52]. A deviation from the normal range of cry-unit durations was associated with asphyxia, meningitis, hydrocephalus, peripheral respiratory distress, central respiratory distress[52], hyperbilirubinemia[57,58], and prenatal exposure to opiate[49] and cocaine[50].

**Pause duration** is defined as the time between the offset of one expiratory utterance and the onset of the next. It is typically during this time that the newborn takes in air and prepares for the next expiration. Together with cry-unit duration, it reflects the neural control of the respiratory system. Past studies show that durational biomarkers are dependent on the state of the infant's respiratory system[52].

## Cry-specific biomarkers extraction

We designed extractors for all cry biomarkers based on their definitions in the clinical literature. Durational biomarkers (cry unit and pauses) are computed in a straightforward way from manually segmented expiration timestamps and further aggregated for each recording using mean, standard deviation, maximum, and minimum statistics.

Most pitch-based biomarkers are extracted from F0 contour computed with an open-source pitch estimator - CREPE[59]. The only exception is dysphonation which relies on spectral flatness computed with Librosa[60].

All pitch-based biomarkers are binary indicators computed for each 10-millisecond audio frame (1 - detected and 0 - otherwise). Hyperphonation is implemented as an indicator of F0 exceeding a pre-defined threshold for a substantial duration. Dysphonation is detected following the same procedure as hyperphonation, except that F0 is substituted by spectral flatness. Vibrato is calculated as an indicator of the oscillating F0 curve detected by measuring the distance between peaks. Glide is implemented as an indicator of a sharp increase or decrease in the F0 sequence. Melody type is represented by five biomarkers that indicate the shape of F0 contour within an expiratory cry unit: falling, rising-falling, rising, falling-rising, or flat.

The thresholds and other parameters of biomarker extractors are tuned using our training database to maximize the classification performance of individual biomarkers. For classification, the indicators of pitch-based biomarkers are aggregated on a per-recording basis as cry unit fraction (number of cry units with biomarker observed in proportion to the total number of cry expiration segments in the recording) and duration fraction (number of frames with biomarker observed in proportion to total number of frames in the recording).

## Selecting a compact subset of biomarkers

First, we calculated Pearson correlation coefficient between each feature and the corresponding positive (Sarnat mild, moderate, severe) or negative label (Sarnat Normal) independently for each site using a subset of train and validation curated birth recordings.

It is hypothesized that the features whose correlation coefficients have the same sign across all three hospitals are reasonably robust discriminators of neuro injury.

Out of 114 total features (88 generic and 26 cry-specific) 18 were selected using this approach. Among these selected features summarized in Extended Data Table 2, 6 come from cry biomarkers while the other 12 are generic voice features.

## Statistical modeling using biomarkers as features

To further study the biomarker features, we conduct classification experiments using logistic regression trained on various feature subsets. Specifically, we consider 88 generic voice features, and 26 cry-specific biomarkers, their combinations and subsets selected based on consistent sign of correlation across three hospitals.

Logistic regression is built using the scikit-learn package[61] with hyperparameters selected using 10-fold cross-validation similar to neural network training setup. The detailed results are summarized in Extended Data Table 3.

By comparing each feature set with its selected counterparts, we conclude that the selected subset of features, although small in size, can achieve classification performance comparable to that achieved by the entire feature set. Although these signal processing-based features are far from neural network performance, they remain valuable due to their interpretability and association with physiological characteristics.

# Data availability

The VGGSound dataset that was used for initial pre-training is publicly available at https://www.robots.ox.ac.uk/~vgg/data/vggsound. An anonymized subset of the Ubenwa unlabelled cry database was released publicly under a Creative Commons Attribution NonCommercial NoDerivatives 4.0 license to encourage machine learning research in infant cry analysis. It was used for adaptation and is available at https://huggingface.co/datasets/Ubenwa/CryCeleb2023. The labeled set of the Ubenwa cry database used for fine-tuning was used under license for the current study and is not publicly available.

## Code availability

We used a proprietary clinical study application developed by Ubenwa Health to record infant cry sounds along with the associated clinical information. The code used for the experiments is closely coupled with proprietary algorithms and tools developed by Ubenwa Health and therefore its release is not feasible. The building blocks are however available as open-source repositories: All machine learning experiments were conducted using python 3.9 (https://www.python.org). Model pre-training and self-supervised cry domain adaptation were done using publicly available SpeechBrain toolkit (https://speechbrain.github.io) and training recipes adapted from related work[24] (https://github.com/zhepeiw/cssl_sound). Model fine-tuning and evaluation were based on commonly used packages including torch 1.13.0, pytorch-lightning 1.8.1, torchaudio 0.13.0, numpy 1.22.4, pandas 1.5.0, hydra 1.2.0 and neptune 0.16.12. Some generic voice features considered in this study were extracted using open-source python package opensmile 2.4.2. We included detailed methods and implementation steps in the Supplementary Methods section to allow for independent replication of the experiments.

## Methods references

32. Simonyan, K. & Zisserman, A. Very Deep Convolutional Networks for Large-Scale Image Recognition. in *ICLR 2015* (eds. Bengio, Y. & LeCun, Y.) (2015).

33. Wang, Z. *et al.* Learning Representations for New Sound Classes With Continual Self-Supervised Learning. *IEEE Signal Process Lett* **29**, 2607–2611 (2022).

34. Park, D. S. *et al.* SpecAugment: A Simple Data Augmentation Method for Automatic Speech Recognition. in *Interspeech 2019* (eds. Kubin, G. & Kacic, Z.) 2613–2617 (ISCA, 2019). doi:10.21437/Interspeech.2019-2680.

35. Spijkervet, J. & Burgoyne, J. A. Contrastive Learning of Musical Representations. in *ISMIR 2021* (eds. Lee, J. H. et al.) 673–681 (2021).

36. Gorin, A. *et al.* Self-Supervised Learning for Infant Cry Analysis. in *ICASSP 2023 - Workshops* 1–5 (IEEE, 2023). doi:10.1109/ICASSPW59220.2023.10193421.

37. Robins, A. V. Catastrophic Forgetting, Rehearsal and Pseudorehearsal. *Connect Sci* **7**, 123–146 (1995).

38. Eyben, F., Wöllmer, M. & Schuller, B. Opensmile: the Munich Versatile and Fast Open-source Audio

Feature Extractor. in *ACM Multimedia Conference* 1459–1462 (ACM, 2010). doi:10.1145/1873951.1874246.

39. Eyben, F. *et al.* The Geneva Minimalistic Acoustic Parameter Set (GeMAPS) for Voice Research and Affective Computing. *IEEE Trans. Affect. Comput.* **7**, 190–202 (2016).

40. Atmaja, B. T. & Akagi, M. On The Differences Between Song and Speech Emotion Recognition: Effect of Feature Sets, Feature Types, and Classifiers. in *IEEE TENCON 2020* 968–972 (2020). doi:10.1109/TENCON50793.2020.9293852.

41. Haider, F., de la Fuente, S. & Luz, S. An Assessment of Paralinguistic Acoustic Features for Detection of Alzheimer's Dementia in Spontaneous Speech. *IEEE J. Sel. Top. Signal Process.* **14**, 272–281 (2020).

42. Li, J.-L., Huang, T.-Y., Chang, C.-M. & Lee, C.-C. A Waveform-Feature Dual Branch Acoustic Embedding Network for Emotion Recognition. *Front. Comput. Sci.* **2**, (2020).

43. Macary, M., Lebourdais, M., Tahon, M., Estève, Y. & Rousseau, A. Multi-corpus Experiment on Continuous Speech Emotion Recognition: Convolution or Recurrence? in *Speech and Computer* (eds. Karpov, A. & Potapova, R.) vol. 12335 304–314 (Springer International Publishing, 2020).

44. Haider, F., Pollak, S., Albert, P. & Luz, S. Emotion recognition in low-resource settings: An evaluation of automatic feature selection methods. *Comput. Speech Lang.* **65**, 101119 (2021).

45. Yang, M. *et al.* Improving Speech Enhancement through Fine-Grained Speech Characteristics. in *Interspeech 2022* 2953–2957 (ISCA, 2022). doi:10.21437/Interspeech.2022-11161.

46. Corwin, M. J., Lester, B. M. & Golub, H. L. The infant cry: What can it tell us? *Curr. Probl. Pediatr.* **26**, 313–334 (1996).

47. Chittora, A. & Patil, H. A. Spectral analysis of infant cries and adult speech. *Int. J. Speech Technol.* **19**, 841–856 (2016).

48. LaGasse, L. L., Neal, A. R. & Lester, B. M. Assessment of infant cry: Acoustic cry analysis and parental perception: Assessment of Infant Cry. *Ment. Retard. Dev. Disabil. Res. Rev.* **11**, 83–93

(2005).

49. Corwin, M. J., Golub, H. L. & Potter, M. Cry Analysis in Infants of Narcotic Addicted Mothers. *Pediatr. Res.* **21**, 180–180 (1987).

50. Lester, B. M. *et al.* The Maternal Lifestyle Study: Effects of Substance Exposure During Pregnancy on Neurodevelopmental Outcome in 1-Month-Old Infants. *Pediatrics* **110**, 1182–1192 (2002).

51. Corwin, M. J. *et al.* Effects of In Utero Cocaine Exposure on Newborn Acoustical Cry Characteristics. *Pediatrics* **89**, 1199–1203 (1992).

52. Golub, H. L. & Corwin, M. J. A Physioacoustic Model of the Infant Cry. in *Infant Crying* (eds. Lester, B. M. & Zachariah Boukydis, C. F.) 59–82 (Springer US, 1985). doi:10.1007/978-1-4613-2381-5_3.

53. Lester, B. M. *et al.* Neurobehavioral Syndromes in Cocaine-Exposed Newborn Infants. *Child Dev.* **62**, 694–705 (1991).

54. Lester, B. M. & Dreher, M. Effects of Marijuana Use during Pregnancy on Newborn Cry. *Child Dev.* **60**, 765–771 (1989).

55. Nugent, J. K., Lester, B. M., Greene, S. M., Wieczorek-Deering, D. & O'Mahony, P. The Effects of Maternal Alcohol Consumption and Cigarette Smoking during Pregnancy on Acoustic Cry Analysis. *Child Dev.* **67**, 1806–1815 (1996).

56. Michelsson, K., Raes, J., Thodén, C.-J. & Wasz-Hockert, O. Sound spectrographic cry analysis in neonatal diagnostics. An evaluative study. *J. Phon.* **10**, 79–88 (1982).

57. Koivisto, M., Wasz-Höckert, O., Vuorenkoski, V., Partanen, T. & Lind, J. Cry Studies in Neonatal Hyperbilirubinemia. *Acta Paediatr.* **59**, 26–27 (1970).

58. Wasz-Höckert, O., Koivisto, M., Vuorenkoski, V., Partanen, T. J. & Lind, J. Spectrographic Analysis of Pain Cry in Hyperbilirubinemia. *Biol. Neonat.* **17**, 260–271 (1971).

59. Kim, J. W., Salamon, J., Li, P. & Bello, J. P. Crepe: A Convolutional Representation for Pitch Estimation. in *ICASSP 2018* 161–165 (2018). doi:10.1109/ICASSP.2018.8461329.

60. McFee, B. *et al.* librosa: 0.10.0. (2023) doi:10.5281/ZENODO.591533.

61. Pedregosa, F. *et al.* Scikit-learn: Machine Learning in Python. *J. Mach. Learn. Res.* **12**, 2825–2830 (2011).

# Acknowledgements

The authors would like to thank all participating patients and hospital staff for helping make the data collection phase of this work possible. We thank J. Lebensold, M. Bellemare and F. Voumard for reviewing the article, and the rest of the Ubenwa team for their support.

# Author contributions

C.C.O. and D.P. contributed to the conception of the study; C.C.O., D.P., S.L., P.O.U., M.A.S, O.A.K., U.E., and D.B. contributed to study design. P.O.U., M.A.S, O.A.K., U.E., G.V. and D.B contributed to acquisition of the data. S.L. contributed to the quality control of the data with support from A.G.. A.G., J.W. and S.L. contributed to curation of data. A.G. contributed to the development of ML models with advice from C.C.O., D.P., and Y.B.. J.W. contributed to development of acoustic biomarkers with advice from C.C.O., A.G. and S.L. A.G., S.L. J.W., C.C.O, D.P., Y.B. contributed to the interpretation of the results; C.C.O, A.G., J.W., and S.L. wrote the manuscript. All authors edited and revised the manuscript.

# Competing interests

The authors declare the following competing interests: C.C.O., S. L., A. G., J. W. have a provisional patent application related to this work in the name of Ubenwa Intelligence Solutions Inc.

# Extended Data

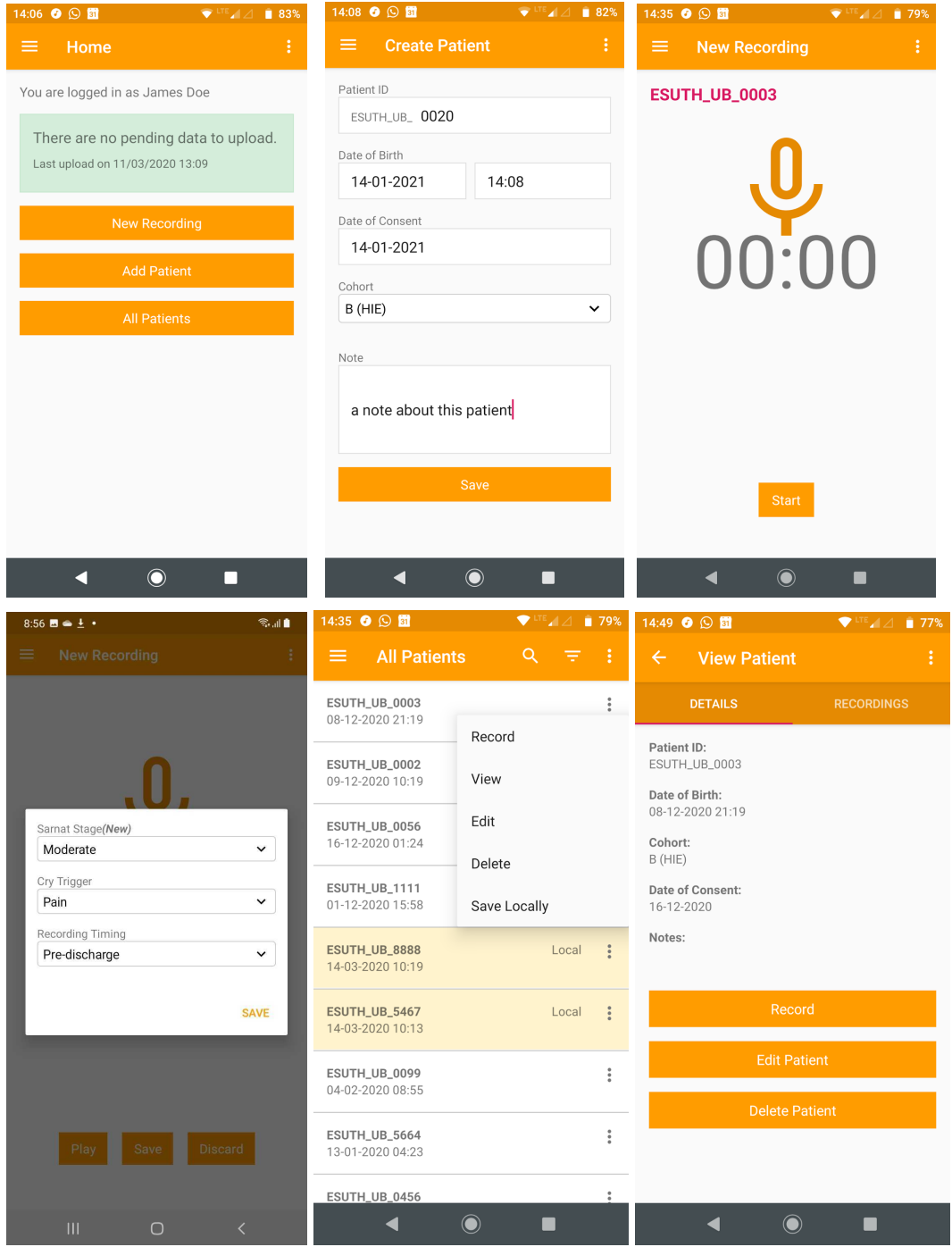

**Extended Data Fig. 1. Data collection interface.** Upon admission, patient information is added to the database along with date of birth and cohort information. Recording is done with Samsung A10 smartphone held at 10-15 cm from the newborn's mouth. One or more recordings were collected at birth and before discharge.

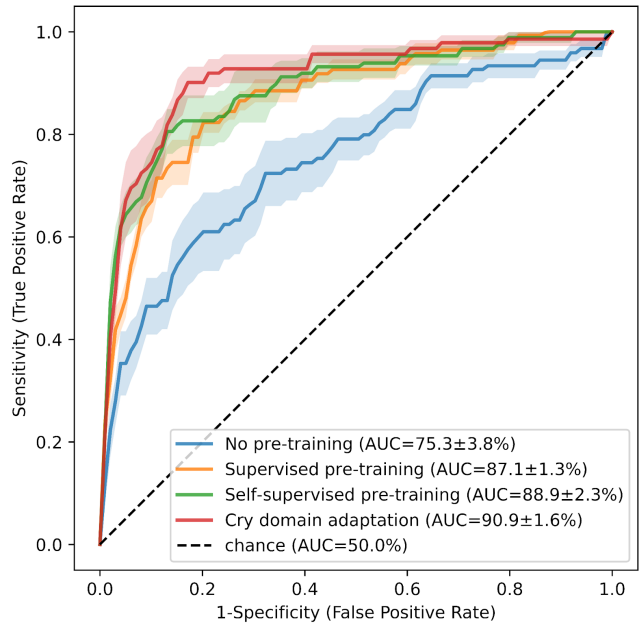

**Extended Data Fig. 2. Performance (ROC) of several models on 10-fold cross-validation.** To tune the hyperparameters of the model (learning rate and fine-tuning strategy), a 10-fold cross-validation approach is used. Training and validation sets are split by 10 folds and for each fold the model is trained with the remaining 9. The ROC curve shows average and standard error intervals of the best-performing hyperparameter sets. Randomly initialized model (without pre-training) performs poorly with only 75.3% AUC. Supervised and self-supervised pre-training on generic audio data (without cry domain adaptation) improves the cross-validation performance significantly. The best cross-validation results are obtained with the model that is further adapted to unlabeled cry sounds.

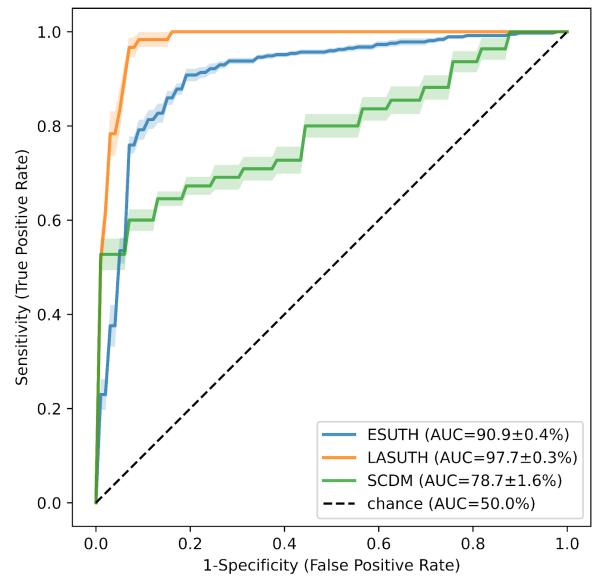

**Extended Data Fig. 3. Test set performance of the best model per site.** The test set performance differs depending on the hospital. With 97.7% AUC for LASUTH, the model obtains only 78.7% on SCDM. This may be attributed to the fact that the model was trained on 50 times more recordings from ESUTH and LASUTH compared to SCDM (see Extended Data Table 1)

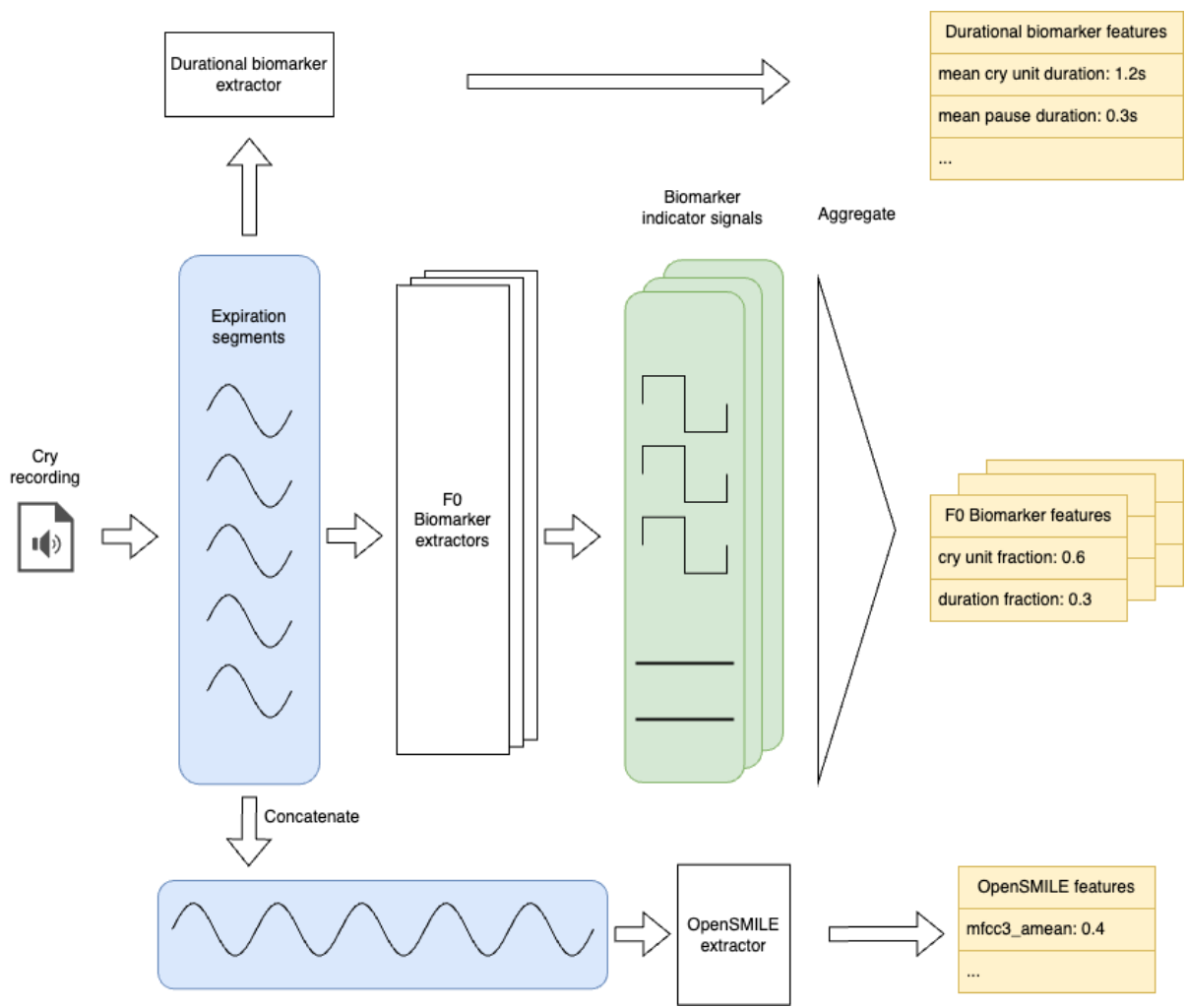

**Extended Data Fig. 4. | Schematic representation of biomarker extraction.** We combine two types of features: cry-specific biomarkers (durational and F0 based, such as cry unit duration, pause duration, hyperphonation, dysphonation, glide, vibrato) and generic audio features extracted using open source tool (OpenSMILE Geneva Minimalist Acoustic Parameter set). Both extractors utilize cry expiration sounds and summarize arbitrary-length recordings with fixed-size feature vectors derived from frame-based signal processing features.

**Extended Data Table 1**. **Characteristics of the dataset used for ML modeling**

|  |  | Train | Val | Test | Total |
|---|---|---|---|---|---|
| Level of encephalopathy (Sarnat score) | Normal | 770 | 55 | 134 | 959 |
|  | Mild | 49 | 7 | 31 | 87 |
|  | Moderate/Severe | 26 | 13 | 23 | 62 |
| Site | ESUTH | 729 | 27 | 84 | 840 |

|        |        |     |     |      |
|--------|-------:|----:|----:|-----:|
| LASUTH | 100    | 35  | 77  | 212  |
| SCDM   | 16     | 13  | 27  | 56   |
| **Total** | **845** | **75** | **188** | **1108** |

**Extended Data Table 2**. **Subset of selected cry-specific biomarkers**

| Biomarker type | Feature name | Correlation with Sarnat |
|---|---|---|
| Cry-specific biomarkers | Fraction of cry units with rising-falling melody | negative |
| | Fraction of cry units with flat melody | positive |
| | Fraction of cry units with glide biomarker | negative |
| | Fraction of frames with glide biomarker | negative |
| | Fraction of cry units with dysphonation | positive |
| | Fraction of frames with dysphonation | positive |
| Generic voice features | slopeUV0-500_sma3nz_amean | negative |
| | slopeV0-500_sma3nz_stddevNorm | positive |
| | slopeV0-500_sma3nz_amean | negative |
| | F2frequency_sma3nz_amean | negative |
| | F3frequency_sma3nz_amean | negative |
| | F3frequency_sma3nz_stddevNorm | positive |
| | mfcc3_sma3_amean | positive |
| | mfcc3V_sma3nz_amean | positive |
| | mfcc3V_sma3nz_stddevNorm | positive |
| | loudness_sma3_stddevFallingSlope | positive |
| | mfcc2_sma3_stddevNorm | negative |
| | mfcc4V_sma3nz_stddevNorm | negative |

**Extended Data Table 3. Logistic regression classification performance using different feature sets**

| Feature set | Number of features | Mean 10-fold Cross-Validation AUC ± standard error | Test AUC % |
|---|---|---|---|
| Voice | 88 | 66.0 ± 2.2 | 59.3 |
| Selected Voice | 12 | 60.9 ± 2.1 | 59.7 |
| Cry | 26 | 65.1 ± 3.5 | 56.9 |
| Selected cry | 6 | 61.9 ± 2.6 | 60.5 |
| Voice + cry | 114 | 68.5 ± 1.8 | 63.7 |
| Selected Voice + selected cry | 18 | 66.1 ± 2.7 | 61.6 |

Six logistic regression models are trained respectively on generic voice features, selected voice features, cry biomarker features, selected cry biomarker features, generic voice features with cry biomarker features, and selected voice features with selected cry biomarker features.